\documentclass{ws-procs9x6}

\begin{document}

\title{Absolute calibration of the LOPES antenna system}

\author{S. NEHLS$\rm^A$, W.~D. APEL$\rm^A$, F. BADEA$\rm^A$, L. B\"AHREN$\rm^B$, K. BEKK$\rm^A$,
        A. BERCUCI$\rm^C$, M. BERTAINA$\rm^D$, P.~L. BIERMANN$\rm^E$, J. BL\"UMER$\rm^{A,F}$,
        H. BOZDOG$\rm^A$, I.~M. BRANCUS$\rm^C$, M. BR\"UGGEMANN$\rm^G$, P. BUCHHOLZ$\rm^G$,
        S. BUITINK$\rm^H$, H. BUTCHER$\rm^B$, A. CHIAVASSA$\rm^D$, K. DAUMILLER$\rm^A$,
        A.~G. DE BRUYN$\rm^B$, C.~M. DE VOS$\rm^B$, F. DI PIERRO$\rm^D$, P. DOLL$\rm^A$,
        R. ENGEL$\rm^A$, H. FALCKE$\rm^{B,E,H}$, H. GEMMEKE$\rm^I$ , P.~L. GHIA$\rm^J$,
        R. GLASSTETTER$\rm^K$, C. GRUPEN$\rm^G$, A. HAKENJOS$\rm{^F}$, A. HAUNGS$\rm{^A}$, 
        D. HECK$\rm^A$, J.~R. H\"ORANDEL$\rm^F$, A. HORNEFFER$\rm^{H,E}$, T. HUEGE$\rm^{A,E}$, 
        K.-H. KAMPERT$\rm^K$, G.~W. KANT$\rm^B$, U. KLEIN$\rm^L$, Y. KOLOTAEV$\rm^G$, 
        Y. KOOPMAN$\rm^B$, O. KR\"OMER$\rm^I$, J. KUIJPERS$\rm^H$, S. LAFEBRE$\rm^H$, 
        G. MAIER$\rm^A$, H.~J. MATHES$\rm^A$, H.~J. MAYER$\rm^A$, J. MILKE$\rm^A$, 
        B. MITRICA$\rm^C$, C. MORELLO$\rm^J$, G. NAVARRA$\rm^D$, A. NIGL$\rm^H$, 
        R. OBENLAND$\rm,^A$, J. OEHLSCHL\"AGER$\rm^A$, S. OSTAPCHENKO$\rm^A$, S. OVER$\rm^G$, 
        H.~J. PEPPING$\rm^B$, M. PETCU$\rm^C$, J. PETROVIC$\rm^H$, T. PIEROG$\rm^A$, 
        S. PLEWNIA$\rm^A$, H. REBEL$\rm^A$, A. RISSE$\rm^M$, M. ROTH$\rm^F$, 
        H. SCHIELER$\rm^A$, G. SCHOONDERBEEK$\rm^B$, O. SIMA$\rm^C$, M. ST\"UMPERT$\rm^F$, 
        G. TOMA$\rm^C$, G.~C. TRINCHERO$\rm^J$, H. ULRICH$\rm^A$, J.~VAN~BUREN$\rm^A$,
        W.~VAN~CAPELLEN$\rm^B$, W. WALKOWIAK$\rm^G$, A. WEINDL$\rm^A$, S. WIJNHOLDS$\rm^B$, 
        J. WOCHELE$\rm^A$, J. ZABIEROWSKI$\rm^M$, J.~A. ZENSUS$\rm^E$, D. ZIMMERMANN$\rm^G$}

\address{$\rm^A$ Institut f\"ur Kernphysik, Forschungszentrum Karlsruhe, Germany\\
         $\rm^B$ ASTRON Dwingeloo, The Netherlands\\
         $\rm^C$ NIPNE Bucharest, Romania\\
         $\rm^D$ Dpt di Fisica Generale dell'Universita Torino, Italy\\
         $\rm^E$ Max-Planck-Institut f\"ur Radioastronomie, Bonn, Germany\\
         $\rm^F$ Institut f\"ur Experimentelle Kernphysik, Uni Karlsruhe, Germany\\
         $\rm^G$ Fachbereich Physik, Universit\"at Siegen, Germany\\
         $\rm^H$ Dpt of Astrophysics, Radboud Uni Nijmegen, The Netherlands\\
         $\rm^I$ IPE, Forschungszentrum Karlsruhe, Germany\\
         $\rm^J$ Ist di Fisica dello Spazio Interplanetario INAF, Torino, Italy\\
         $\rm^K$ Fachbereich Physik, Uni Wuppertal, Germany\\
         $\rm^L$ Radioastronomisches Institut der Uni Bonn, Germany\\
         $\rm^M$ Soltan Institute for Nuclear Studies, Lodz, Poland}

\maketitle

\abstracts{Radio emission in extensive air showers arises from an
  interaction with the geomagnetic field and is subject of theoretical
  studies.  This radio emission has advantages for the detection of
  high energy cosmic rays compared to secondary particle or
  fluorescence measurement methods. Radio antennas like the LOPES30
  antenna system are suited to investigate this emission process by
  detecting the radio pulses. The characteristic observable parameters
  like electric field strength and pulse length require a calibration
  which was done with a reference radio source resulting in an
  amplification factor representing the system behavior in the
  environment of the KASCADE-Grande experiment. Knowing the
  amplification factor and the gain of the LOPES antennas LOPES30 is
  calibrated absolutely for systematic analyses of the radio
  emission.}

\section{Introduction}
The long known radio emission in extensive cosmic ray air showers
(EAS) is again under investigation with new fully digital radio
antennas. Nearly 40 years ago, in the early 1960's this nano-second
short weak pulses in EAS were detected and basically confirmed with
theoretical predictions. With recent theoretical studies (Huege and
Falcke\cite{hueg05}), using a more detailed Monte-Carlo technique, and
a new generation of radio telescopes the comparison of predictions and
measured radio emission in EAS provides us with a capable method for
EAS investigation. The stochastic production process of EAS is a
complicated phenomenon. Therefore as many observables as possible are
needed to reconstruct the primary shower parameters correctly. The
digital radio antenna field of LOPES30 placed inside the existing
multiple detector-component experiment KASCADE-Grande\cite{nava03} is
now calibrated absolutely allowing us to measure precisely the long
known radio emission in EAS and their dependencies on primary shower
parameters like arrival direction, primary particle mass and energy.

\section{Radio emission in EAS}
Analytical calculations in the early 1970's\cite{alla71} of the expected
electric field strength $\epsilon_{\nu}$, the lateral distribution of
$\epsilon_{\nu}$ and the dependence on the shower direction predict
electric field strengths at ground level in the range of
$\epsilon_{\nu}\approx5$ -- $15~\mu$V/m/MHz ($E_{\omega}\approx0.5$ --
$2.5~\mu$V/m/MHz) for primary energies $\sim10^{17}$~eV. The
definition of the quantity $E_{\omega}$ and a conversion factor for
$\epsilon_{\nu}$ can be found in \cite{hueg03}. On the basis of the so
called geosynchroton-effect a new analytical model for the calculation
of the electric field strength $E_{\omega}$ was developed 2003 by
Huege and Falcke. The results of the simulations have been summarized
with a parametrisation formula to get expected electric field strength
$E$ occurring in EAS.  From this parametrisation formula one gets
electric field strength at ground in the range of $E_{\omega}\approx3$
-- $5~\mu$V/m/MHz also for primary energies $\sim10^{17}$~eV. There
has never been a common agreement about the absolute field
strength\cite{atra78} and the values cited decreased over time to a
tenth of a $\mu$V/m/MHz. For the absolute calibration of LOPES30 these
values give a first benchmark to our detection thresholds but they do
not really represent typical values of electrical field strengths
occurring in model prediction of EAS.

\section{Calibration setup for LOPES30}
\begin{figure}[hb]
\centerline{\epsfxsize=3.9in\epsfbox{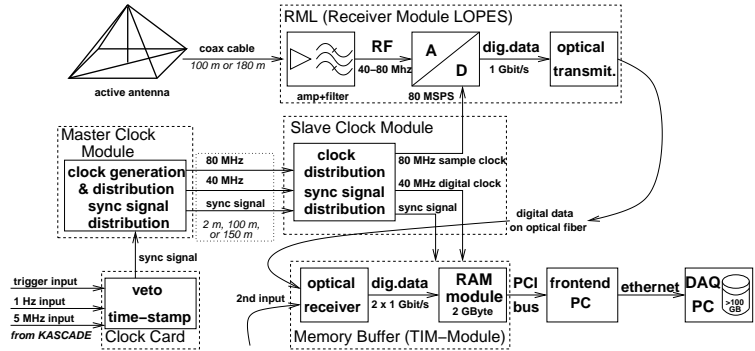}}   
\caption{Scheme of LOPES30 electronic. Incoming radio pulses from 
  EAS were detected with inverted-V-shaped antennas, transmitted over
  100m to 180m coax cable to the Receiver Module (RML), digitizes, and
  stored.\label{fig:scheme}}
\end{figure}

With LOPES10 the ``proof of principle'' in detecting radio emission
from EAS was achieved by comparing relative field strengths in the
antenna array \cite{falc05} and comparing them with the parameters
obtained by KASCADE-Grande. The analysis was done without a precise
absolute calibration and therefore only a qualitative comparison with
theoretical predictions was possible. An absolute comparison can be
done by knowing the system (Fig.~\ref{fig:scheme}) response to a
calibrated well defined signal where the knowledge of the voltage
amplitude and signal phase is included.
Due to their inverted-V-shape the antennas are most sensitive to
vertical EAS and less sensitive to highly inclined signals
($>70^{\circ}$) except around 60~MHz as shown in
figure~\ref{fig:antenna21}, left. A mechanically needed quadratic
ground plate (2.5 x 2.5~m$^2$) of aluminum modifies the antenna gain,
i.e.  it increases the antenna gain in the range of 60~MHz towards
highly inclined signals and decreases the gain for vertical signals.
From a commercial reference radio source (VSQ 1000\cite{scha05}) the
electrical field strength $E$ in a certain distance is known and the
emitted time-continuous and frequency discrete signal is used. This
means that the reference radio source emits in 1~MHz, 5~MHz, or 10~MHz
steps a defined sine wave, e.g. at 55~MHz around four orders of
magnitude higher in power than EAS radio emission, at 10~m distance in
the main direction. In our calibration setup the radio source was
placed $\approx10~m$ above the top of the LOPES antennas. Mounted at
the end of a wooden beam fixed on an extension arm of a crane we
determined for each antenna an individual frequency dependent
amplification factor. These values represent the overall system
behavior to the input signal emitted by the reference antenna and
therefore all active and passive components in the electronic system
(see figure~\ref{fig:scheme}) contribute with their individual gain.
It is more difficult to calculate an amplification factor from a
single component calibration of the full electronical chain, because
some components do not have exactly 50~$\Omega$ impedance. For the
calibration the transmitted power $P_t$, the gain $G_t$ of the
reference radio source, and the gain $G_r$ of the LOPES antenna
correlate with the received power $P_r$:
\begin{equation}
P_r=\left(\frac{\lambda}{4d\pi}\right)^2G_rG_tP_tcos^2(\beta)
\end{equation}
In the temporary setup of a merely simulated LOPES antenna gain $G_r$
the received power $P_r$ can be determined. The polarization angle
$\beta$ is needed to take into account that the LOPES antennas are
linearly polarized and therefore the angle between polarization axes
of the emitter and polarization axes of the detecting antenna modifies
the received power. For all LOPES antennas we succeeded to measure the
received power in the main sensitivity direction.

\section{Results}
\begin{figure}[t]
\centerline{\epsfxsize=1.75in\epsfbox{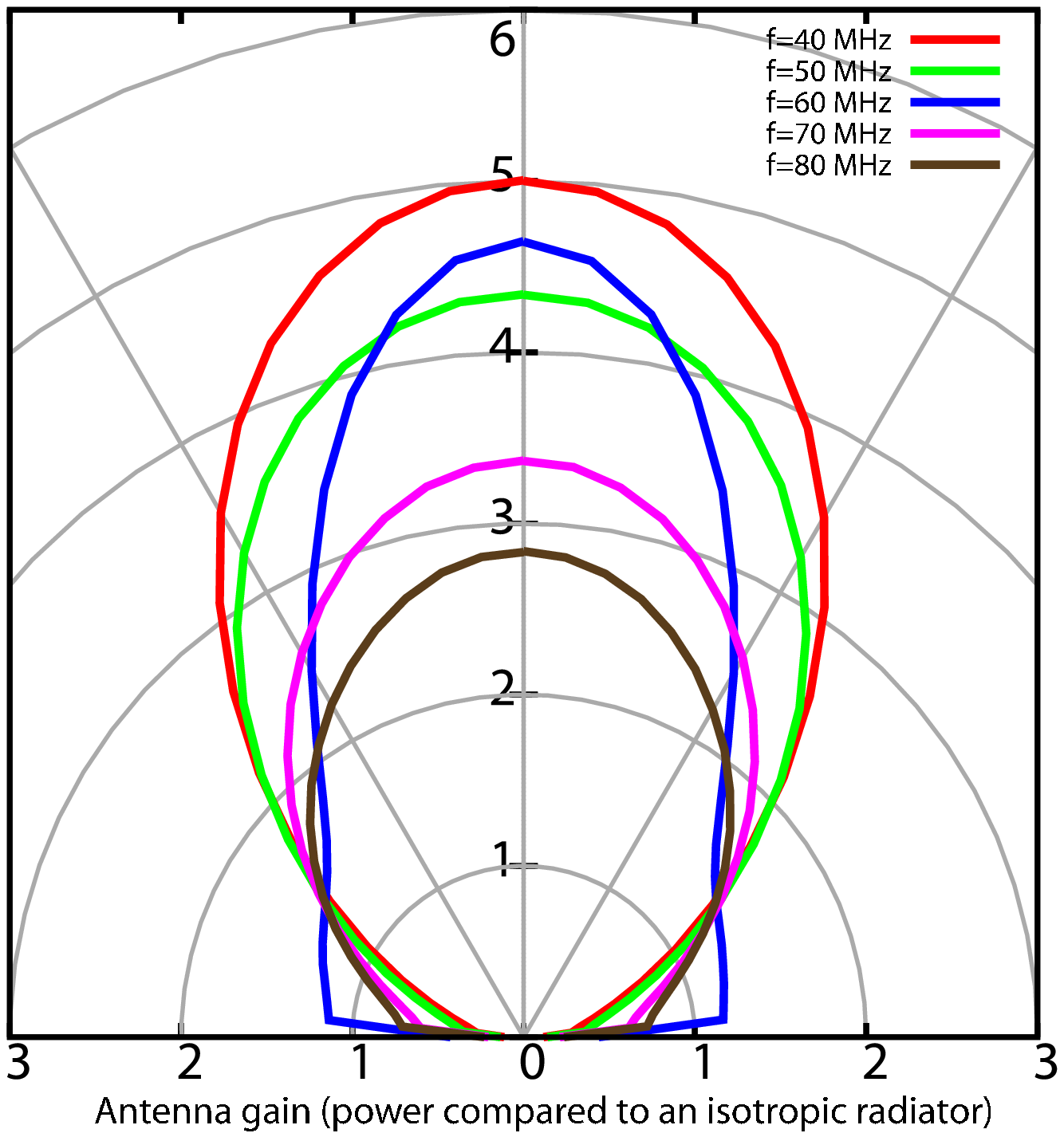}
            \epsfxsize=2.5in\epsfbox{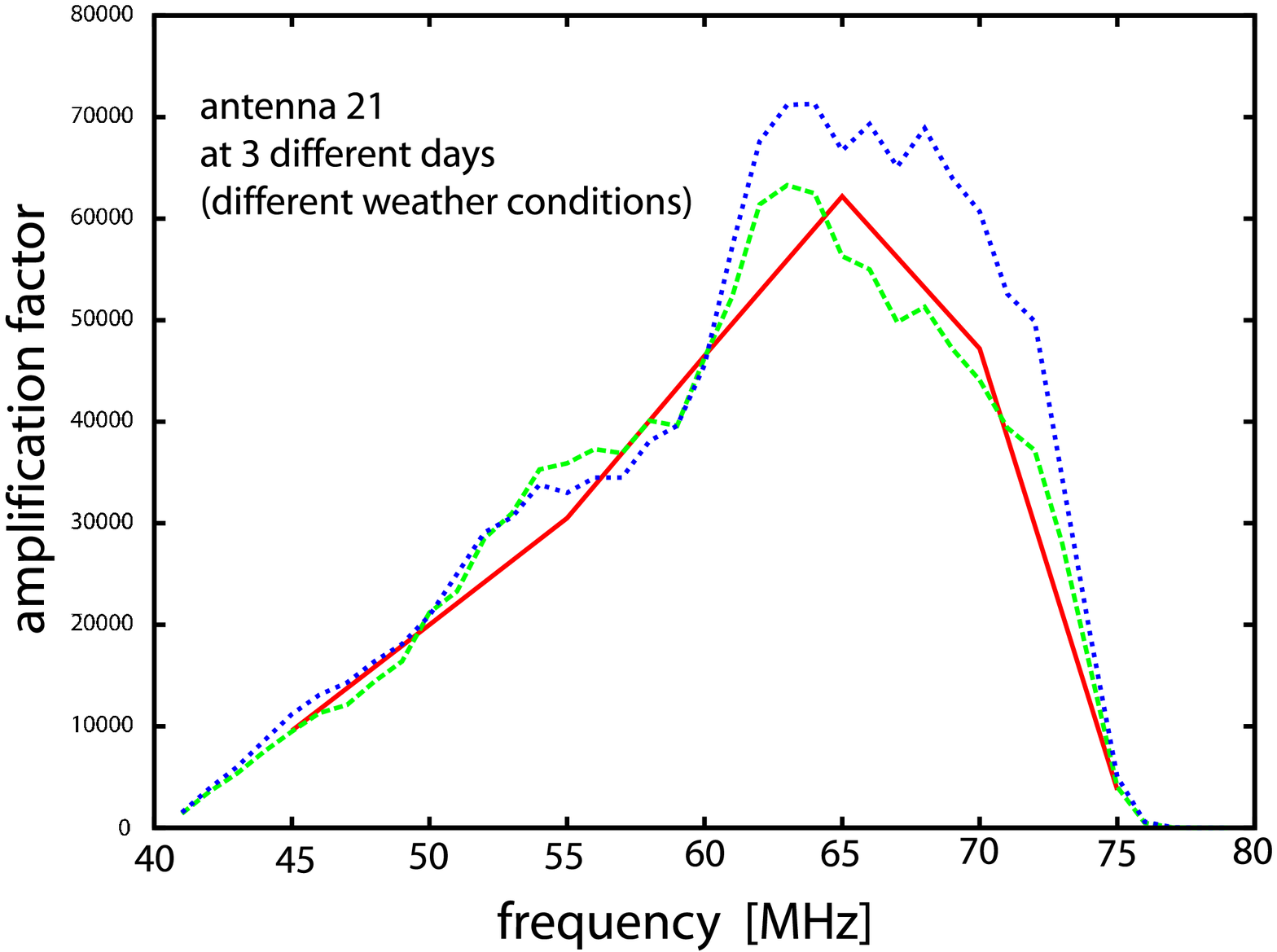}}   
\caption{Left: Antenna gain from 40 to 80~MHz in a polar diagram 
  (simulation). Right: Estimated amplification factors for one
  antenna. From 40~MHz to 80~MHz the influence of different conditions
  can be seen, especially above 60~MHz one can see deviations in the
  order of 25~\%. Lower solid curve in 5~MHz steps for dry conditions.
  Middle dotted curve in 1~MHz steps wet conditions. Upper dotted
  curve in 1~MHz steps at rain fall.\label{fig:antenna21}}
\end{figure}

In a campaign of three days the measurements were done in the 5MHz or
1MHz step mode of the reference radio source. The fraction of
transmitted power $P_t$ to received power $P_r$ is proportional to an
amplification factor calculated from a 9.8~msec dataset for each
LOPES antenna. We determined such amplification factors for all
antennas which can vary from antenna to antenna by a factor of ten
with a typical uncertainty of around 15\%. The relatively large factor
of ten between the antennas is mostly caused by the characteristics of
the bandpass filter and is one of the important contributions to the
amplification factor. The uncertainties at 50~MHz are larger compared
to frequency ranges above and below because of amateur radio
communication occurring in this band. For one antenna we measured the
received power in different weather conditions. In
figure~\ref{fig:antenna21} three curves are shown, representing the
amplification factors as a function of frequency, for very dry
conditions, wet conditions, and also during rain fall.
As a first result it is obvious that the conditions during the
calibration measurement such as soil humidity, rain fall, or relative
humidity influenced the values of the amplification factor. These
first results need further detailed investigations. A weather
dependent correction factor for the LOPES antenna system can minimize
these calibration uncertainties. Furthermore above 60~MHz the
variations in the amplification factor are much stronger than below,
and there is no significant connection with the polarization axis
between the two antennas or a systematic shift of the curves relative
to each other. To eliminate this influence periodic calibration
campaigns are needed to better understand the performance of the LOPES
antenna system.

\end{document}